\documentclass[10pt, journal,twocolumn]{IEEEtran}
\IEEEoverridecommandlockouts
\usepackage{cite}
\usepackage{amsmath,amssymb,amsfonts}
\usepackage{graphicx,setspace}
\usepackage{textcomp}
\usepackage{xcolor}
\usepackage{subfigure}
\usepackage{algorithm,algpseudocode}
\usepackage{pdfpages}
\usepackage{epstopdf}

\newcommand\CONDITION[2]%
  {\begin{tabular}[t]{@{}l@{}l@{}}
     #1&#2
   \end{tabular}%
  }
\algdef{SE}[WHILE]{While}{EndWhile}[1]%
  {\algorithmicwhile\ \CONDITION{#1}{\ \algorithmicdo}}%
  {\algorithmicend\ \algorithmicwhile}
\algdef{SE}[FOR]{For}{EndFor}[1]%
  {\algorithmicfor\ \CONDITION{#1}{\ \algorithmicdo}}%
  {\algorithmicend\ \algorithmicfor}
\algdef{S}[FOR]{ForAll}[1]%
  {\algorithmicforall\ \CONDITION{#1}{\ \algorithmicdo}}
\algdef{SE}[REPEAT]{Repeat}{Until}{\algorithmicrepeat}[1]%
  {\algorithmicuntil\ \CONDITION{#1}{}}
\algdef{SE}[IF]{If}{EndIf}[1]%
  {\algorithmicif\ \CONDITION{#1}{\ \algorithmicthen}}%
  {\algorithmicend\ \algorithmicif}%
\algdef{C}[IF]{IF}{ElsIf}[1]%
  {\algorithmicelse\ \algorithmicif\ \CONDITION{#1}{\ \algorithmicthen}}
\usepackage{textcomp}
\usepackage{array}
\usepackage{tabularx}
\usepackage{booktabs}
\usepackage{mathtools}
\usepackage{comment}
\usepackage{amsthm}
\usepackage{url}
\usepackage{siunitx}

\usepackage{subfig}

\usepackage{epstopdf}

\DeclarePairedDelimiter{\norm}{\lVert}{\rVert}

\allowdisplaybreaks

\begin{document}
\title{RL-Based Cargo-UAV Trajectory Planning and Cell Association for Minimum Handoffs, Disconnectivity, and Energy Consumption
\thanks{This 
work is funded in part by the Natural 
Sciences and Engineering Council Canada (NSERC).}}
\author{\IEEEauthorblockN{  Nesrine~Cherif\IEEEauthorrefmark{1}, Wael Jaafar\IEEEauthorrefmark{2}, Halim Yanikomeroglu\IEEEauthorrefmark{3}, and Abbas Yongacoglu\IEEEauthorrefmark{1}. \\
	\IEEEauthorblockA{\IEEEauthorrefmark{1}School of Electrical Engineering and Computer Science, University of Ottawa, Ottawa, ON, Canada\\
\IEEEauthorrefmark{2}Department of Software and IT Engineering, École de Technologie Supérieure (ÉTS), Montreal, QC, Canada\\		
  \IEEEauthorrefmark{3}Department of Systems and Computer Engineering, Carleton University, Ottawa, ON, Canada.	
		}
		}
}

\maketitle
\linespread{1.1}
\begin{abstract}
Unmanned aerial vehicle (UAV) is a promising technology for last-mile cargo delivery. However, the limited on-board battery capacity, cellular unreliability, and frequent handoffs in the airspace are the main obstacles to unleash its full potential. Given that existing cellular networks were primarily designed to service ground users, re-utilizing the same architecture for highly mobile aerial users, e.g., cargo-UAVs, is deemed challenging. Indeed, to ensure a safe delivery using cargo-UAVs, it is crucial to 
utilize the available energy efficiently, while guaranteeing reliable connectivity for command-and-control and avoiding frequent handoff. To achieve this goal, we propose a novel approach for joint cargo-UAV trajectory planning and cell association. 
Specifically, we
formulate the cargo-UAV mission as a multi-objective problem aiming to 1) minimize energy consumption, 2) reduce handoff events, and 3) guarantee cellular reliability along the trajectory. We leverage 
reinforcement learning (RL) to jointly optimize the cargo-UAV's trajectory and cell association. Simulation results demonstrate a performance improvement of our proposed method, in terms of handoffs, disconnectivity, and energy consumption, compared to benchmarks.

\end{abstract}

\vspace{-15pt}
\section{Introduction}
\textcolor{black}{Cargo unmanned aerial vehicles (cargo-UAVs) are beginning to grab a portion of the worldwide delivery market over many industries due to their increased flexibility and mobility}. For instance, Amazon Prime Air service is under development for beyond visual line-of-sight (BVLoS) operations. 
In August 2021, Amazon got the approval of the federal aviation administration (FAA) to operate its UAV fleet.
With this milestone, cargo-UAV based delivery is expected to increase in the near future. Hence, cargo-UAV safety and security are paramount for the roll-out of aerial delivery and its public acceptance \cite{3daerialhighway}. \textcolor{black}{To ensure a safe operation at BVLoS, reliable and permanent command and control (C\&C) communications are crucial for cargo-UAVs \cite{fotouhiSurvey}.
Specifically, cellular networks are required to be highly available, reliable, and secure to satisfy 
C\&C requirements. The viability and feasibility of BVLoS cellular-connected UAVs have been examined in several works \cite{Ozger2018,Li2019,Amorim2020}.}
However, re-utilizing existing cellular infrastructures for aerial operations brings challenges such as unreliable radio links due for instance to down-tilted antennas of terrestrial base stations (BSs) \cite{cherifdownlink}. Besides, the mobility inheritance of cargo-UAVs may lead to frequent switching from one serving cell to another, a.k.a. handoff, hence, resulting in more signaling overhead that degrades the radio link's quality.

To guarantee cellular link reliability and reduce latency in terrestrial networks, authors of \cite{alkhateeb2018machine} proposed a deep learning-based handoff prediction strategy for millimiter wave communications.
For cellular-connected UAVs, several works focused on optimizing UAV trajectories for BVLoS operations. For instance, in \cite{fakhreddine2019handover}, the authors set up a testbed for UAV missions. The performance of such a system was evaluated in \cite{euler2019mobility} in terms of radio link failure and handoff rate.
Moreover, the authors of \cite{chen2020efficient} proposed a reinforcement learning (RL) based algorithm that balances between handoff rate and radio link quality for a pre-known UAV trajectory. Finally, we investigated for the first time in \cite{CherifICC} cargo-UAV trajectory planning with consideration for energy consumption, communication link disconnectivity, and handoffs. We proposed dynamic programming-based path planning and studied the impact of parameters such as disconnectivity \textcolor{black}{and handoff rates}, and \textcolor{black}{impact of BSs'} distribution \textcolor{black}{on} energy consumption.

\textcolor{black}{To the best of our knowledge, only \cite{CherifICC} examined cargo-UAV trajectory planning from the viewpoints of energy consumption, cellular connection dependability, and handoff, together.
However, the suggested solution there neglected the issue of cell affiliation and presupposed that the cargo-UAV is always linked with the cell with the strongest signal, which may result in frequent handoffs.}
Consequently, we propose in this work a novel algorithm for joint trajectory planning and cell association that aims to minimize energy consumption and number of handoff events leveraging RL, while guaranteeing a reliable communication link with the terrestrial network. \textcolor{black}{It is noteworthy that the reliability of a communication link can be evaluated in terms of cellular disconnectivity time during the cargo-UAV mission.}

The remaining of the paper is organized as follows. Section II presents the system model. In Section III, we formulate the related problem. Section IV details the proposed RL-based trajectory planning and cell association algorithm for cargo-UAVs, while simulation results are discussed in Section V. Finally, Section VI concludes the paper.

\section{System Model}
\label{sec:sysmodel}
We consider a 3-dimensional (3D) geographical area in the airspace of the city of Leuven, Belgium, centered at coordinates 50°52'45.3"N 4°42'08.1"E, based on a Cartesian coordinates system $(x,y,z)$.  \textcolor{black}{It is important to highlight that the data herein was generated using the same real data simulator mentioned in \cite{colpaert2018aerial}}. 
We assume that a cargo-UAV is deployed to deliver an item from a retailer warehouse with coordinates $\textbf{q}_w=(x_w,y_w,0)$  to a preset drop-off location $\textbf{q}_d=(x_d,y_d,0)$ where the customer is expected to pick it up, e.g., at a community drop-off location or the private customer's porch \cite{3daerialhighway}. The cargo-UAV is expected only to travel either vertically, i.e., at fixed $(x,y)$, or horizontally, i.e., at fixed altitude $z=h$.
To safely execute its mission, the cargo-UAV relies on cellular connectivity of the existing terrestrial network for C\&C exchanges with a central unit on the ground. 

\subsection{Channel Model}
The air-to-ground (A2G) communication channel model has been defined by the 3rd Generation Partnership Project (3GPP). 
It encompasses the BS's antenna gain, LoS probability, and pathloss expressions. 
\textcolor{black}{For further details on the adopted A2G channel model and its components, we kindly refer the reader to \cite[Section II-A]{CherifICC}.
}

\subsection{Received Power Expression}

\begin{figure*}[t]
	\centering
	\includegraphics [trim={0.1cm 1cm 0.2cm 0.1cm},clip,width = 0.7\linewidth]{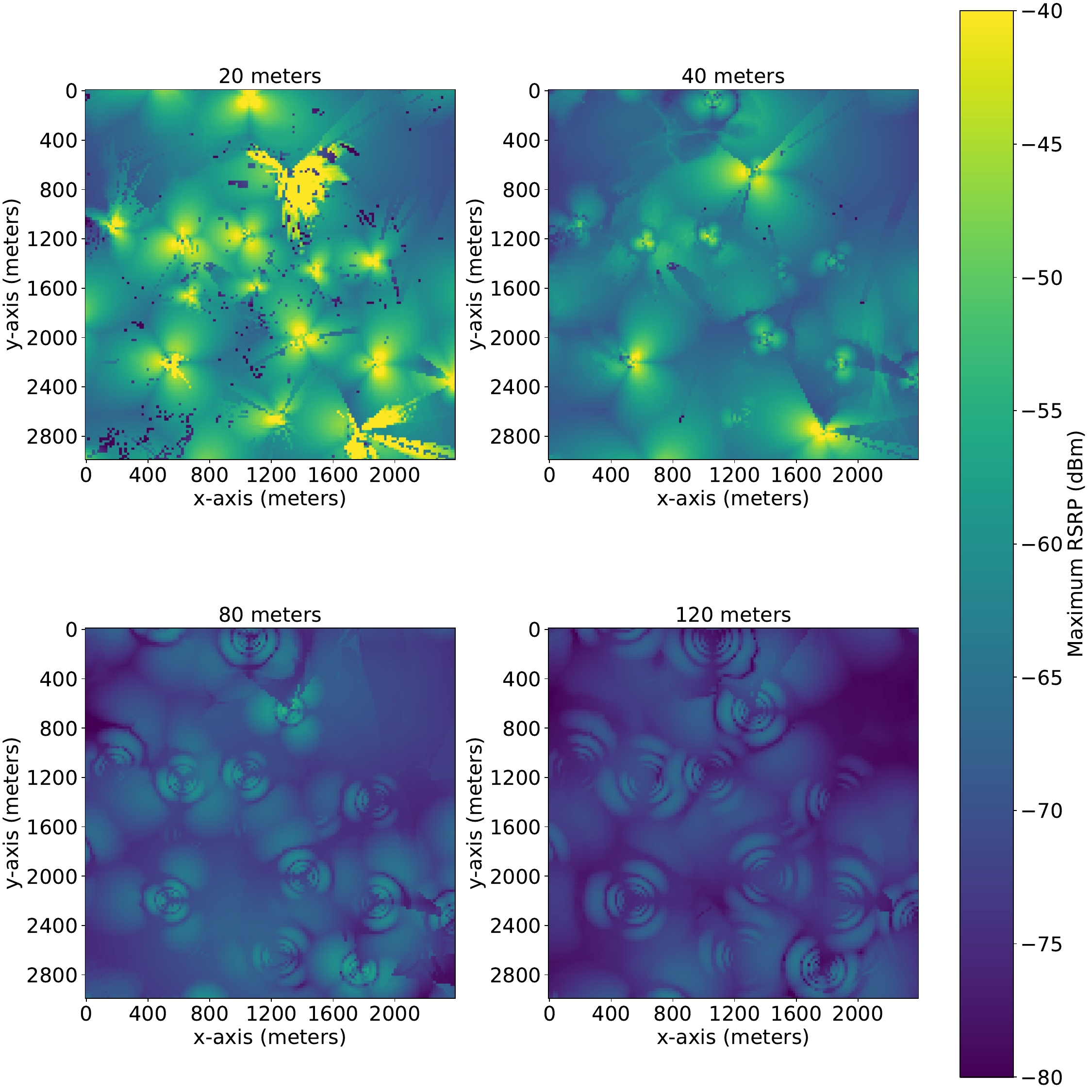}
	\caption{Maximum RSRP heatmap of a part of the city of Leuven, Belgium, for different cargo-UAV altitudes (Area is $3 \times 3$ km$^2$ and  $f_c=1.8$ GHz) \cite{colpaert2018aerial}.}
	\label{fig:heatmap}
\end{figure*}

Based on the channel model and assuming that the cargo-UAV is equipped with an omni-directional antenna with unit gain, the average received power from the $i^{th}$ cell, i.e., the Reference Signal Received Power (RSRP), is given by 
\begin{equation}
    \label{eq:Pr}
    P_{r,i}=P_T + G_i-L_i, \; \forall i=1,\ldots,M,
\end{equation}
where $P_T$ is the transmit power of any ground cell in dBm. $G_i$ is the array radiation pattern of the $i^{th}$ cell given by \cite[eq.(5)]{CherifICC}, and $L_i$ is the average pathloss of the channel between the cargo-UAV and the $i^{th}$ cell highlighted in \cite[eq.(7)]{CherifICC}. Finally, $M$ is the total number of cells in the area.
For the sake of simplicity, we assume that frequency planning is \textcolor{black}{orthogonal and} optimal, thus eliminating inter-cell
interference at cargo-UAVs. 
Accordingly, the reliability of cellular connectivity depends on the level of RSRP at any potential location in the airspace. For illustrative purposes, we present in Fig. \ref{fig:heatmap} the maximum RSRP level at each grid point of the $(x,y)$ plane at different flying altitudes $h \in \{20, 40, 80, 120 \}$ meters. At low altitudes, e.g., $h=20$ meters, scattered coverage holes are more frequent than for $h=40$ meters. This is due to the presence of more blockages such as buildings, trees, etc., in the Leuven urban area. \textcolor{black}{In contrast, at elevated altitudes, high pathloss coupled with the down-tilt of terrestrial-BS' antennas weaken extensively the signal strength, thus resulting in more coverage holes areas, as shown for instance at $h=120$ meters.}


Leveraging the data presented in Fig. \ref{fig:heatmap}, the cargo-UAV is considered disconnected when, at its actual location, no cell can guarantee an RSRP equal or higher than a pre-defined RSRP threshold, denoted $T_{\rm th}$\textcolor{black}{\footnote{\textcolor{black}{Here, the cargo-UAV operates in a noise-limited environment where the RSRP sensitivity $T_{\rm th}$ translates into a minimum required achievable data rate.}}}. Moreover, a handoff event occurs when the cargo-UAV is associated with a different cell than the previous one to which it was connected. During a cargo-UAV mission, it is \textcolor{black}{important} to trigger \textcolor{black}{seldom} handoff events, as the latter would destabilize the C\&C link and generate an additional amount of overhead data.

\subsection{Cargo-UAV Energy Consumption Model}
The cargo-UAV energy consumption depends on its size, weight, and power, along its movement and communication regimes \cite{jaafar2020dynamics}. It consists of the propulsion energy and the communication energy. 
Typically, communication energy is very small compared to the propulsion energy, and thus, can be neglected. Using a rotary-wing cargo-UAV, the propulsion energy to fly for a distance $d$ is
\begin{equation}
\small
    \label{eq:ener}
    E(v,d)={P_{\rm prop}}(v) \; d/{v}, \; v>0,
\end{equation}
where $P_{\rm prop}(v) $ is the propulsion power given by \cite[eq.(10)]{CherifICC} and $v$ is the cargo-UAV velocity.
Let $E_C$ and $E_S$
be the initial cargo-UAV energy capacity and emergency reserve energy, respectively.
$E_C$ is available for the mission, while $E_S$ is used for emergency pull-backs. Hence, the cargo-UAV's available energy to travel between the departure and arrival locations is
\begin{equation}
\small
    E_A(v,h)=E_C-E_S-2 E(v,h),
\end{equation}
where $E(v,h)$ is the required energy to take-off from the dockstation or to land on the drop-off location. 

\vspace{-10pt}

\section{Problem Formulation}

The cargo-UAV is required to deliver a package from the retailer warehouse (at coordinates $\textbf{q}_{w}$) to a consumer drop-off location (at coordinates $\textbf{q}_{d}$).  
Our objective is to achieve this mission while consuming minimum energy, triggering minimum handoff events, while keeping a reliable communication.

Let $\mathcal{T}=\{\textbf{q}_1,\ldots, \textbf{q}_{|\mathcal{T}|}\}$ be the trajectory of the cargo-UAV, defined by the locations $\textbf{q}_i=(x_i,y_i,h)$. Also, let $\mathcal{C}=\{c_1,\ldots, c_{|\mathcal{T}|}\}$ be the set of cell IDs associated to the cargo-UAV locations in $\mathcal{T}$, where $c_i \in \{1,\ldots, M\}$. It is noteworthy that at any grid location in the 2D plane at altitude $h$, the cargo-UAV receives RSRP measurement reports from all potential serving cells. Finally, we assume that the set of experienced RSRPs along the trajectory are represented by $\mathcal{R}=\{r_1,\ldots, r_{|\mathcal{T}|}\}$. Hence, the joint trajectory planning and cell association problem can be formulated as follows:
\begin{subequations}
	\begin{align}
	\label{eq:P1}
	{\rm P1:}\min_{\mathcal{T},\mathcal{C}}  &\quad w_{\rm en} \sum_{i=1}^{|\mathcal{T}|-1} E(v,d_{i,i+1})+ w_{\rm sig} \sum_{i=1}^{|\mathcal{T}|-1} \mathcal{D}(d_{i,i+1})\nonumber\\
   	&+\quad w_{\rm ho}\sum_{i=1}^{|\mathcal{T}|-1}\eta_{i,i+1} \nonumber  \tag{P1}
	\\
	\label{c2_21} \text{s.t.}\quad & \sum_{i=1}^{|\mathcal{T}|-1} E(v,d_{i,i+1}) \leq E_{A}(v,h),  \nonumber \tag{P1.a}\\
	\label{c2_2} &\textbf{q}_1 = \textbf{q}_{w}, \text{ and }\textbf{q}_{|\mathcal{T}|} = \textbf{q}_{d}, \tag{P1.b}
	\end{align}
\end{subequations}
where $w_{\rm x}$, $\rm x \in \{\rm en,\ sig,\ ho\}$ are the weights \textcolor{black}{describing the importance given to each metric,} energy consumption, radio link disconnectivity and handoffs, respectively, such that $w_{\rm en}+w_{\rm sig}+w_{\rm ho}=1$. We have $d_{i,i+1}=\norm{\textbf{q}_{i}-\textbf{q}_{i+1}}$, and $\mathcal{D}(d_{i,i+1})=d_{i,i+1}$ if $\textcolor{black}{P_{r,i}}<T_{\rm th}$ and $\textcolor{black}{P_{r,i+1}} < T_{\rm th}$; and $\mathcal{D}(d_{i,i+1})=0$ otherwise. \textcolor{black}{$\mathcal{D}(d_{i,i+1})$ indicates a travelled distance in a coverage hole, i.e., without connectivity.
Finally variable $\eta_{i,i+1}$ reflects the handoff event, i.e., $\eta_{i,i+1}=1$ if $c_i \neq c_{i+1}$ and $\eta_{i,i+1}=0$ otherwise. 
}

\textcolor{black}{A capacitated vehicle routing problem (CVRP) can be used to describe problem (P1) \cite{Lenstra1981}. Indeed, the CVRP is defined as choosing a vehicle's route with the goal of reducing its overall transport expenses. Logically, the cargo-UAV and its costs in terms of energy consumption, disconnectivity, and handoffs in our system, may be assimilated to the CVRP's vehicle and transportation costs, respectively.
Given that the CVRP is known to be an NP-hard problem \cite{Lenstra1981}, then, by restriction, problem (P1) is also NP-hard.}
\vspace{-10pt}
\section{Proposed RL-based Trajectory Planning and Cell Association}
To solve the cargo-UAV constrained trajectory planning and cell association problem, we propose here an RL-based approach. 
First, we briefly overview the adopted RL method. Then, the framework for the proposed RL-based trajectory planning and cell association algorithm, including the definition of states, actions, and rewards are detailed.

\vspace{-10pt}
\subsection{RL Background}
In RL, an agent, i.e., the cargo-UAV, interacts with its environment, i.e., the airspace. Specifically, it executes an action given its current state and the anticipated future reward. At an instant $t$, the agent in state $s$ takes an action $a$. Based on $(s,a)$, the agent receives a reward $\rho$. The process is iteratively repeated to maximize the expected reward over time and reach the end of the mission. The RL algorithm can be presented as a Markov decision process (MDP) defined by the $(\mathcal{S},\mathcal{A},\mathcal{P}_{sa},\beta,\mathcal{W})$ where the set of states is encompassed in $\mathcal{S}$, $\mathcal{A}$ is the set of actions. Moreover, $\mathcal{P}_{sa}$ is the state transition probability of a state $s\in \mathcal{S}$ and an action $a\in \mathcal{A}$. $\beta \in [0,1)$ is the discount factor. Finally, the reward function of a pair of state-action is $\mathcal{W}$. 
The optimal policy of the MDP aims to maximize the sum of discounted rewards by associating an optimal action to each state. We adopt here Q-learning to learn the optimal policy for the MDP\textcolor{black}{\footnote{\textcolor{black}{Note that more recent and sophisticated RL methods than Q-learning can be explored to solve the developed problem. Such an effort is left within the scope of a future work. }}}.  
The optimal policy adopted by the agent is the set of actions per state with the highest Q-values. The Q-value is given by the Bellman equation as
\begin{equation}
\small
    \label{eq:bellman}
    {Q}_{t+1}(s,a)= (1-\alpha){Q}_{t}(s,a)+\alpha\left[\mathcal{W}_{t+1}+\beta \; \underset{a\prime\in \mathcal{A}}{\max}{Q}_{t}(s',a') \right],
\end{equation}
where $Q_t(s,a)$ is the Q-value for $(s,a)$ pair at episode (or time instant) $t$, $\alpha$ is the learning rate, $\mathcal{W}_{t+1}$ is the reward of episode $t+1$, and $Q_t(s',a')$ is the Q-value for state-action $(s',a')$ in episode $t$.


\subsection{Proposed  RL-based Solution}

We present here the proposed RL-based trajectory planning and cell association approach to solve problem (P1). The RL agent hosted by the cargo-UAV
periodically collects its state. Then, it determines the best action, in terms of motion and cell association, and performs it. The state,
action and reward of the RL agent are defined as follows:

Cargo-UAV's \textit{state} is defined as $s=\{\textbf{q}_s,c_s\}$ where $\textbf{q}_s=(x_s,y_s,h)$ is the 3D location of the cargo-UAV in the airspace and $c_s \in \mathcal{C}$ is the corresponding serving cell ID. In state $s$, an action $a$ is drawn from action space $\mathcal{A}$, which includes pairs of potential motion direction, i.e., east (E), west (W), norht (N), south (S), north-east (NE), north-west (NW), south-east (SE), and south-west (SW), and cell IDs. The \textit{reward function} is designed from the objective function of (P1), such that energy consumption, handoffs, and disconnectivity are minimized. Specifically, we express the reward function as 
\begin{equation}
\small
\label{eq:reward}
    \mathcal{W}(s,a)= w_{\rm en}W_{\rm mo}(s,a)+w_{\rm sig}W_{\rm sig}(s,a)+w_{\rm ho}W_{\rm ho}(s,a),
\end{equation}
where 
\begin{align}
\small
    W_{\rm mo}=\left\{ 
  \begin{array}{ c l }
    -1/\sqrt{2}  & \textrm{if } \text{motion direction}\in \{\rm N, S, E, W\}  \\
    -1                 & \textrm{otherwise},
  \end{array} 
\right.
\end{align} 
is the UAV motion (i.e., energy consumption) penalty, 
\begin{align}
\small
    W_{\rm ho}=\left\{ 
  \begin{array}{ c l }
    0 &  \textrm{if } c_{s_t} = c_{s_{t+1}}  \\
   \textcolor{black}{-1}              &  \textrm{otherwise},
  \end{array}
\right.
\end{align} is the handoff cost, and the disconnectivity cost is given by
\begin{align}
\small
    W_{\rm sig}(s,a)=\left\{ 
  \begin{array}{ c l }
    0 &  \textrm{if RSRP}_{c_{s_{t+1}}}\geq T_{\rm th}  \\
    \textcolor{black}{-1}             &  \textrm{otherwise}.
  \end{array}
\right.
\end{align}

\textcolor{black}{It is imperative to underscore that the weights employed in eq.(6), $w_{\rm x}$, $\rm x \in \{\rm en,\ sig,\ ho\}$, with respect to each optimizing metric, namely, RSRP level, energy consumption, and handoff rate, primarily serve to elucidate the latter's importance in the path planning of the cargo-UAV mission. For instance, the RSRP metric, which delineates the cellular link reliability, is expected to have the highest weight value in instances of disconnectivity, given that such occurrences are regarded as the most undesirable events, such as when the cargo-UAV loses connection and as a result it jeopardizes the safety of properties and individuals on the ground. Conversely, handoff events and, to a lesser extent, trajectory detours, are anticipated to receive lower weights compared to disconnectivity events.} The RL agent should be trained for a sufficient number of episodes to understand the environment dynamics and improve its decision-making, thus reaching the destination in the most efficient way. The proposed RL-based algorithm is presented in Algorithm \ref{algo:rl}. Specifically, the RL agent begins with no knowledge about the environment, i.e., an empty Q-table for all state-action combinations. The action space $\mathcal{A}$ encompasses all combinations of the motion directions and cell IDs at any given grid location of the 2D airspace plane. Then, the Q-table is built through the exploration/exploitation of the environment, making random/optimized actions, and calculating the corresponding reward. The latter is used to update the Q-value of the related state-action pair using (\ref{eq:bellman}). 
These steps are repeated until the destination is reached within an episode. The execution of this algorithm for a sufficient number of episodes ensures convergence towards the optimal policy. The output of Algorithm \ref{algo:rl} is the best strategy of trajectory and associated cells during the delivery mission.

\begin{algorithm}[t]
\caption{Proposed RL-based Solution}
\label{algo:rl}
\small
\begin{algorithmic}[1]
\Require{Matrix $\mathbf{H}$ of size $(J\times J \times M')$ is the cellular connectivity heatmap at altitude $h$ 
where $J$ is the airspace grid length and width and $M'<M$ is the number of cells with the highest RSRPs; 
step size; 
cargo-UAV speed $v$; RSRP threshold $T_{\rm th}$; Available energy $E_A$; starting location $\textbf{q}_w$; arrival location $\textbf{q}_d$.}
\Ensure{Optimal policy; Trajectory $\mathcal{T}$; Serving cells $\mathcal{C}$.}
\State Initialize Q-table matrix $\textbf{Q}$ with  size $(J\times J \times 8 \times M')$ with $8 \times M'$ is the size of the action space $\mathcal{A}$.
\While {maximum number of episodes is not reached}
\State  Initialize $\mathcal{T}=\{\mathbf{q}_1\}$ and $\mathcal{C}=\{c_1\}$.
\While {destination is not reached }
\State Generate $\kappa \in [0,1]$ 
\If {$\kappa<\epsilon$}
\State Pick randomly an action from $\mathcal{A}$.
\Else
\State Pick $a^*=\underset{a\in \mathcal{A}}{\rm argmax}{Q}(s,a)$.
\EndIf
\State compute the reward $\mathcal{W}(s,a)$ using (\ref{eq:reward}).
\State Update the Q-value for $(s,a)$ using (\ref{eq:bellman}).
\State $\epsilon=\max\{\epsilon_{\min},\epsilon-\epsilon_{\rm dec}\}$ with $\epsilon_{\min}$ is the minimum allowed value for $\epsilon$ and $\epsilon_{\rm dec}$ is the decay for each taken action.
\EndWhile
\EndWhile
\State Return 
best $(\mathcal{T}^*,\mathcal{C}^*)$. 
\end{algorithmic}
\end{algorithm}

\vspace{-10pt}
\section{Simulation Results}
In this section, we assess the performance of the RL-based cargo-UAV trajectory planning with respect to cellular service reliability and handoffs. Hence, we compare our proposed algorithm to other benchmarks, i.e., shortest path and RSRP-aware path planning. \textcolor{black}{The onboard battery's capacity is the sole factor that significantly restricts the calculation of the shortest path baseline. As a result, the cargo-UAV sets the standard for low battery usage by taking the shortest route while being always connected to the BS with the strongest RSRP along the route.} In contrast, the RSRP-aware RL-based trajectory planning algorithm determines the path that increases cellular reliability, i.e., with the lowest cellular disconnectivity time while ignoring handoff events. 

We assume a 3 $\times$ 3 km$^2$ area in the city of Leuven, Belgium, where 20 three-sector BSs and 2 two-sector BSs are deployed. Each sector accounts for a cell, with a total number of 64. The area is divided into grid cells with a step of $20$ meters on both X and Y axis. We set the following parameters as $v=\ 30$ m/s and $E_C-E_S=2500$ kJ. Finally, typical values of the cargo-UAV related parameters are taken from \cite[Table I]{zeng2019}.

In any given 3D location (at a fixed altitude $h$), the received RSRP from each available cell in the area is measured and stored as part of the environment states space.
For the benchmark algorithms, only the cell with the strongest RSRP is considered to evaluate cellular disconnectivity and handoff performances, whereas our proposed algorithm considers in its states the first three strongest RSRPs, i.e., $N=3$. 
Unless stated otherwise, the cellular service RSRP threshold is set to $T_{\rm th} = -65\ $ dBm, \textcolor{black}{$w_{\rm en}=2.5\%$, $w_{\rm sig}=90\%$, and $w_{\rm ho}=7.5\%$}. For the RL parameters, we set $\beta=0.9$, $\alpha=0.01$, $\epsilon_{\min}=0.01$, and $\epsilon=0.99$. The total number of training episodes is fixed to $10^4$.


\begin{figure}[t]
	\centering
	\includegraphics[trim={0.5cm 0.2cm 1.5cm 1cm},clip,width=0.85\linewidth]{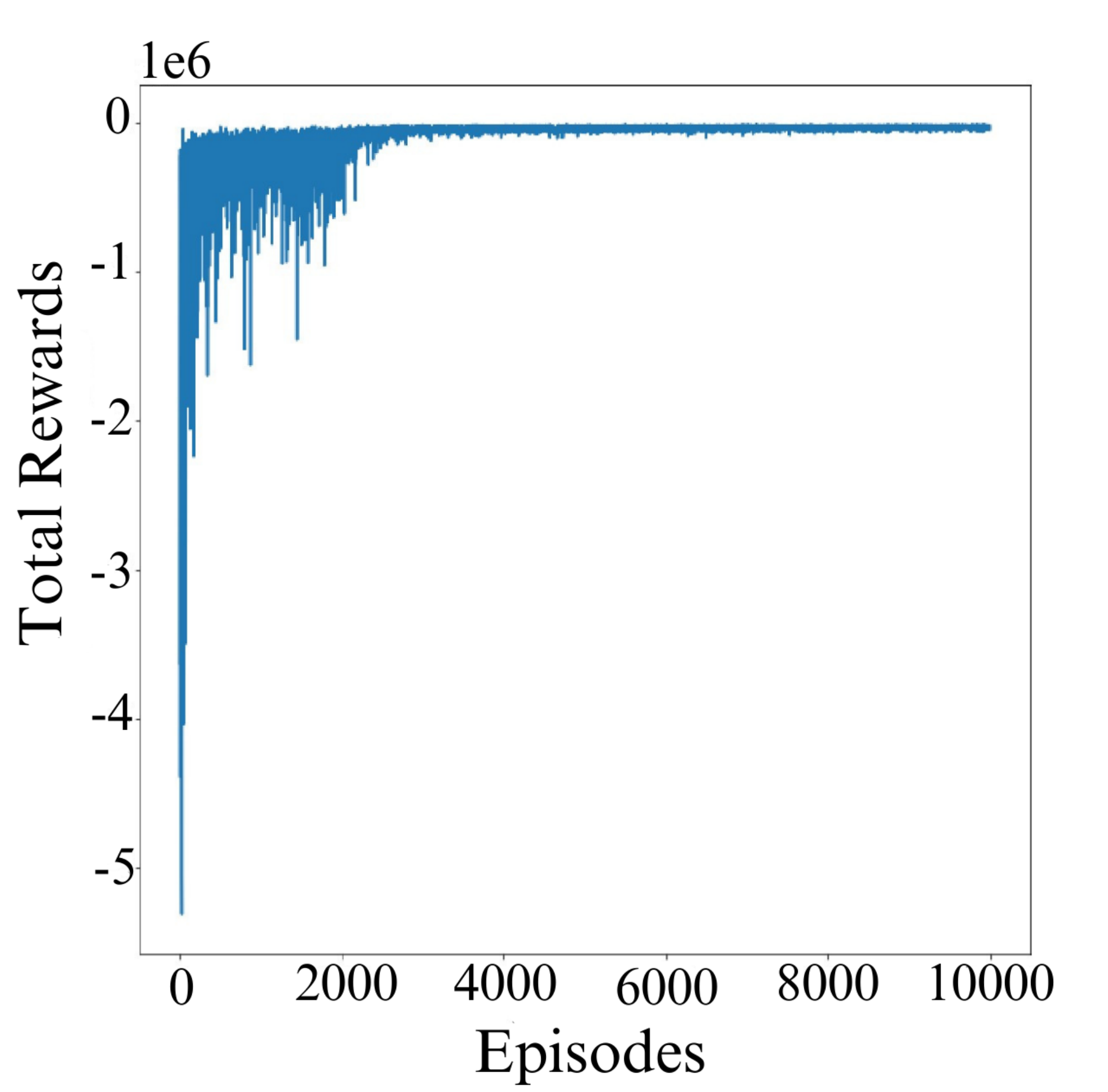}
	\caption{Training convergence of proposed algorithm. 
	}
	\label{fig:learn}
\end{figure}
We show in Fig. \ref{fig:learn} the learning curve of the proposed RL-based path planning algorithm. First, the cargo-UAV starts exploring the airspace without any prior knowledge and by randomizing its action, and accordingly constructs its Q-table by associating each realized action to a prospect reward. As the number of training episodes increases, the exploration phase is given less importance and the cargo-UAV starts to exploit the acquired experience to advance towards its destination. This translates into a proportional reward increase with training. As it can be seen, our algorithm converges from episode 3500, and the reward is maximized close to the maximum value 0.

\begin{figure}[t]
\centering
	\subfigure[Number of handoffs.]{
		\includegraphics[trim={0.5cm 5.5cm 1.5cm 6.2cm},clip,width=0.8\linewidth]{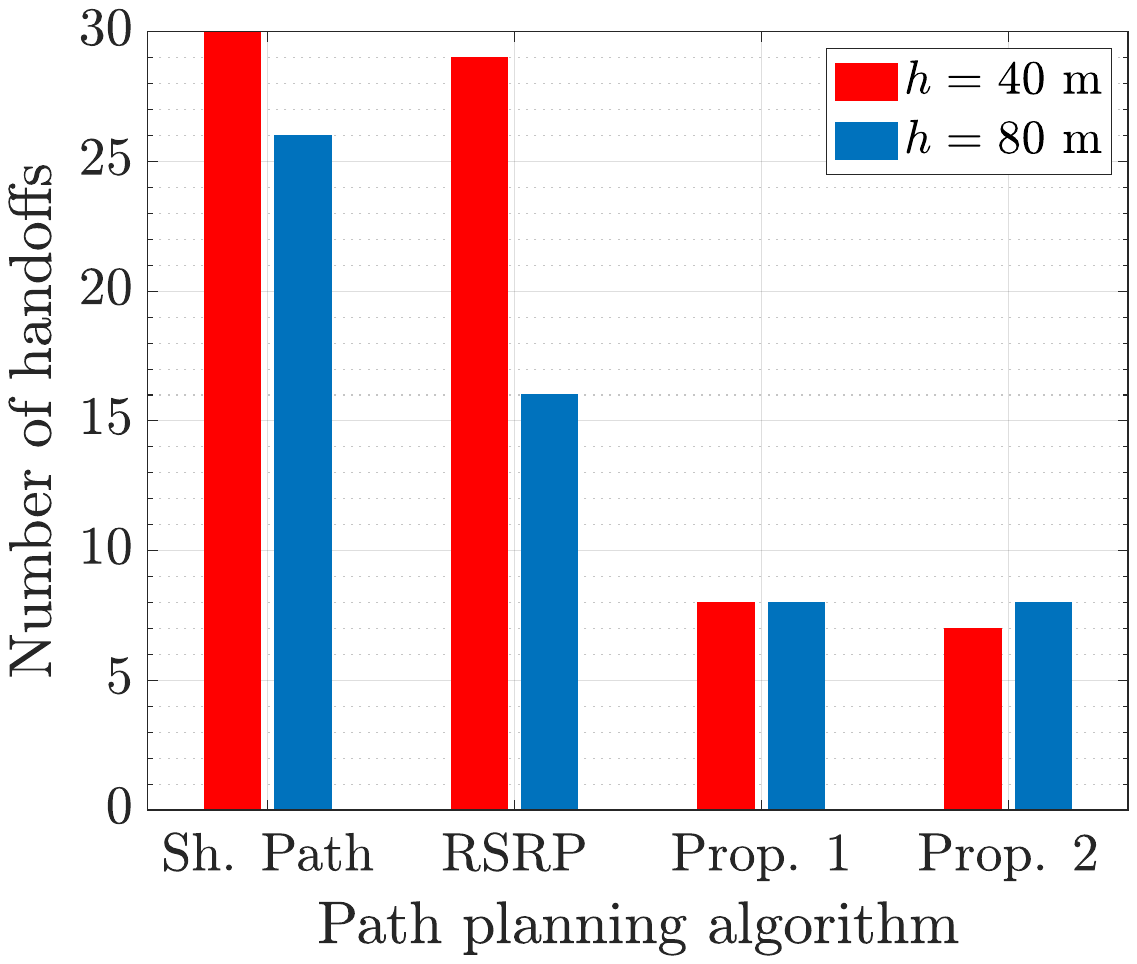}}
	\vfill
	\subfigure[Disconnectivity along cargo-UAV mission (\%).]{ 	
	\includegraphics[trim={0.5cm 5.5cm 1.5cm 6.2cm},clip,width=0.8\linewidth]{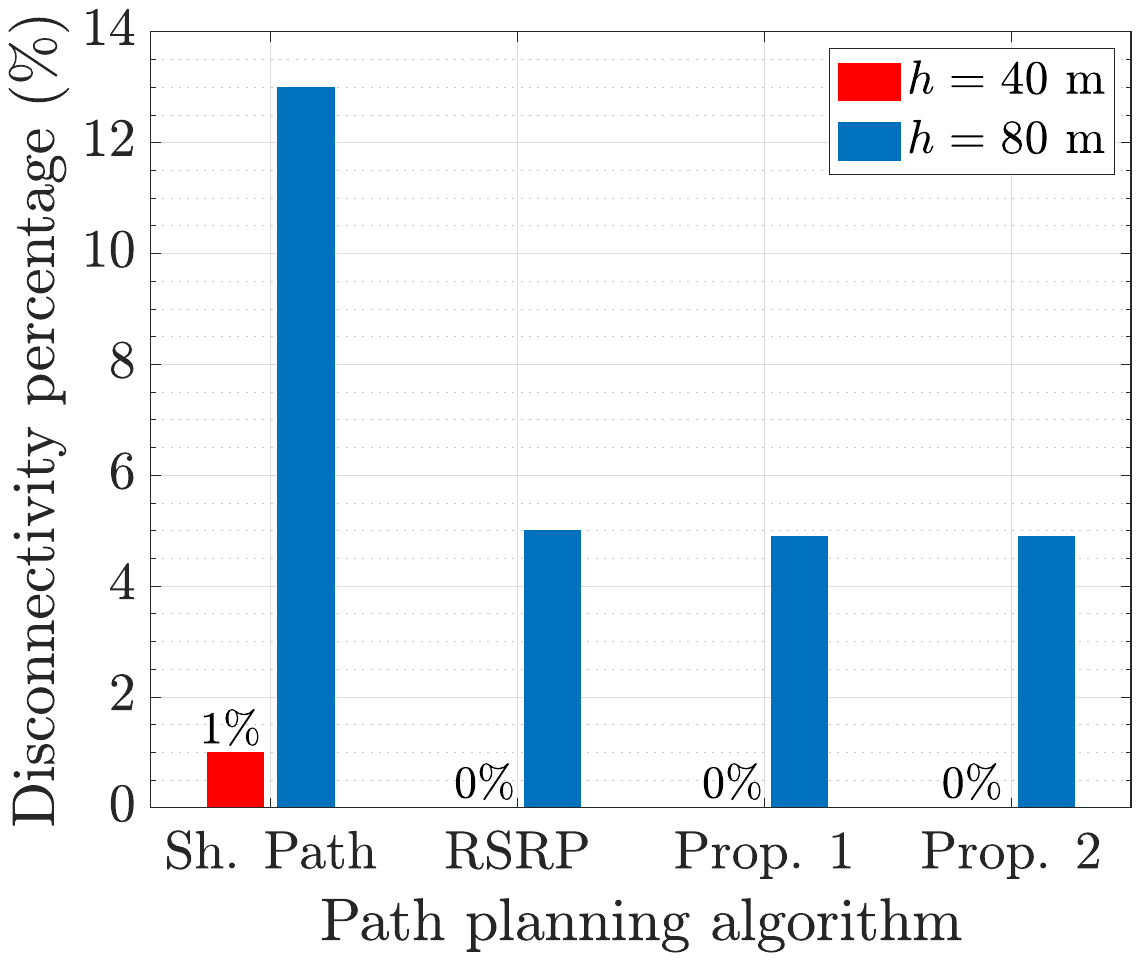}
	}
	\vfill
	\subfigure[Battery consumption (\%).]{ 
	\includegraphics[trim={0.5cm 5.5cm 1.5cm 6.2cm},clip,width=0.8\linewidth]{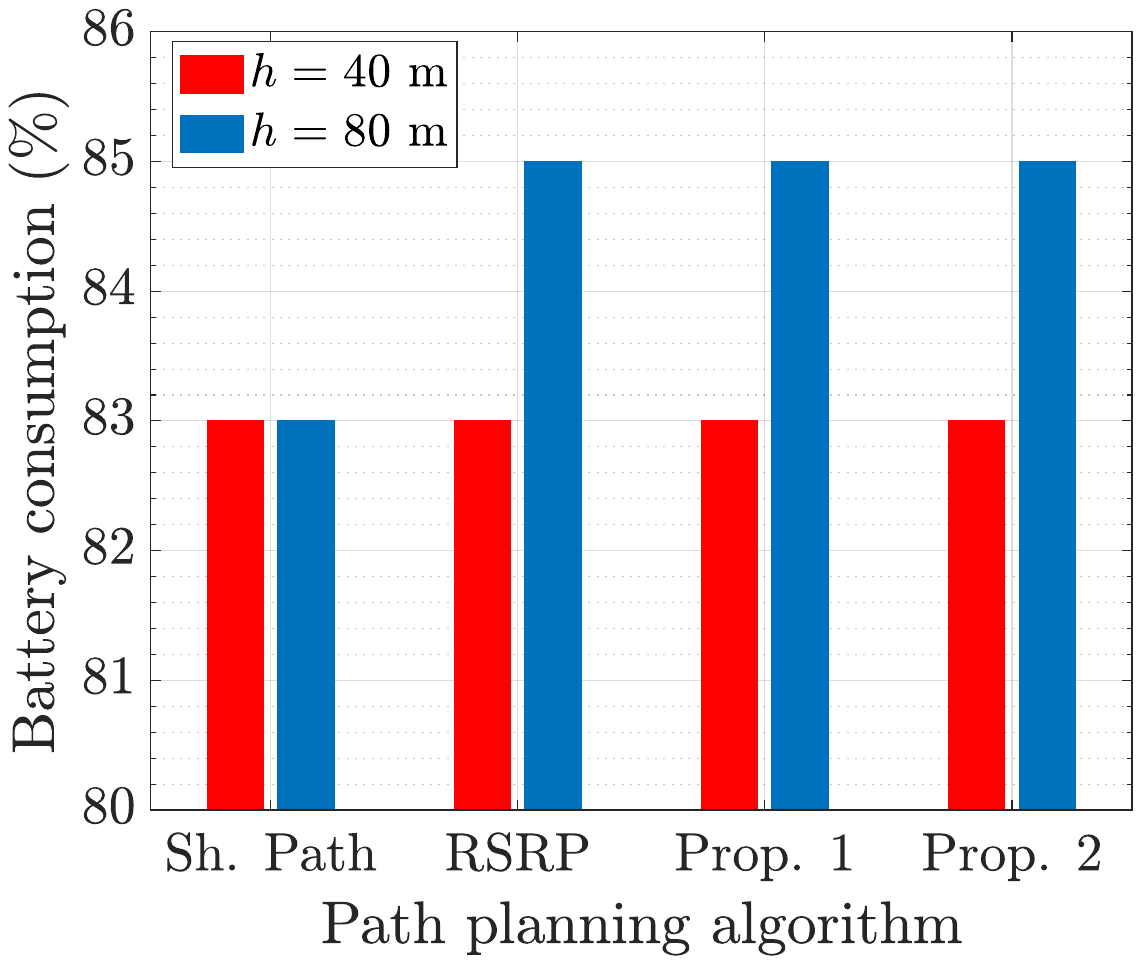}
	}
	\caption{Comparison of path planning algorithms.}
	\label{fig:detail}
\end{figure}

\begin{figure}[t]
	\centering
	\includegraphics[trim={0cm 7cm 0.1cm 6cm},clip,width=0.9\linewidth]{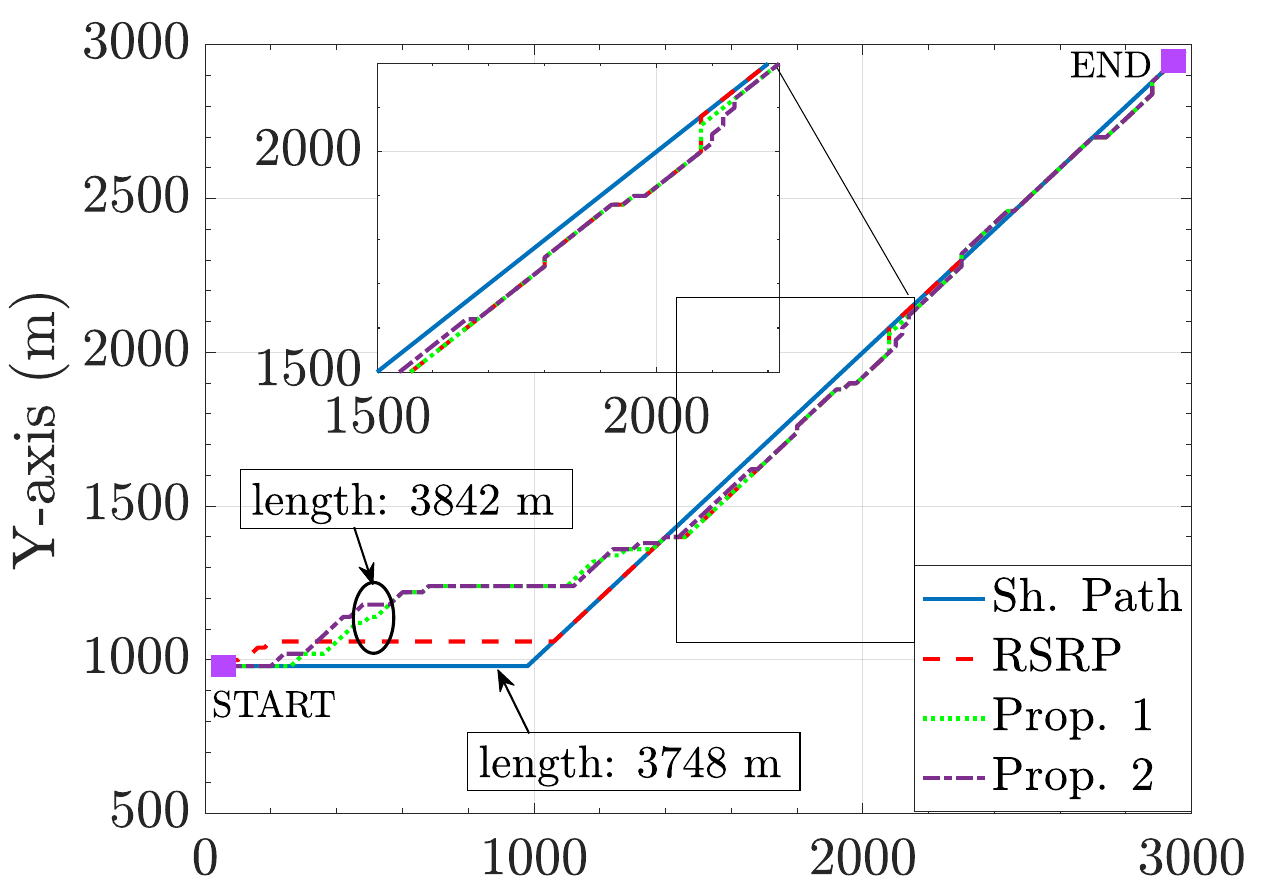}
	\caption{Cargo-UAV trajectories ($h=80$ m). }
	\label{fig:uavroute}
\end{figure}

Fig. \ref{fig:detail} compares the performance of the proposed algorithm against those of benchmarks, i.e., \textcolor{black}{``Sh. Path'' (for shortest path)}
and \textcolor{black}{ ``RSRP'' (for only RSRP-aware)} \cite{CherifICC} algorithms, in terms of number of handoffs, battery consumption percentage, and disconnectivity percentage along the mission's trajectory. For the proposed algorithm, we considered two configurations, the first with \textcolor{black}{$(w_{\rm en},w_{\rm sig},w_{\rm ho})=(2.5\%,90\%,7.5\%)$}, \textcolor{black}{called ``Prop. 1''} and the second with \textcolor{black}{$(w_{\rm en},w_{\rm sig},w_{\rm ho})=(4\%,80\%,16\%)$}, \textcolor{black}{named ``Prop. 2''}.
According to Figs. \ref{fig:detail}.a-\ref{fig:detail}.b, the shortest path algorithm experiences the highest number of handoffs and disconnectivity percentage for any flying altitude $h$. This is expected since the shortest path aim to reach the destination in the fastest possible way, regardless of other factors. As a consequence, it consumes the least energy as shown in Fig. \ref{fig:detail}.c.
For RSRP-Aware algorithm, the number of handoffs and disconnectivity percentage are lower than for the previous method, since the cargo-UAV adjusts its trajectory to keep RSRP above the sensitivity threshold, whenever possible. \textcolor{black}{To achieve that, the cargo-UAV traveled extra distances to avoid coverage holes, especially at $h=80$ m where the RSRP signals are weaker. This has resulted in a higher battery consumption than for the shortest path benchmark, i.e., an additional 2\% battery drainage.}  
For the proposed algorithm, the number of handoffs and disconnectivity percentage are the lowest, while the consumed energy is approximately the same as the shortest path at $h=40$ m and as the RSRP-Aware one at $h=80$ m. 
\textcolor{black}{In fact, the energy used by the extra detours of the proposed method, which were taken to reduce both handoffs and disconnectivity (shown in Fig. \ref{fig:uavroute}), compensates for the energy consumption of handoff signalling method, thus assuring better handoff and disconnectivity performances than the other methods.} Also, when the handoffs weight $w_{\rm ho}$ \textcolor{black}{is increased in the reward function from 7.5\% to 16\%}, the number of handoffs can be further reduced, in particular at lower altitudes.

In Fig. \ref{fig:uavroute}, 
we present the resulting trajectories of the Fig. \ref{fig:detail}'s algorithms.
Although the trajectories of our proposed approach (\textcolor{black}{green and purple} lines) experience several detours, it still achieves the mission with the best performances in terms of number of handoffs and disconnectivity, at the cost of a slight increase in energy consumption.

\begin{figure}%
    \centering
   \includegraphics[trim={1.6cm 6cm 2cm 7cm},clip,width=0.9\linewidth]{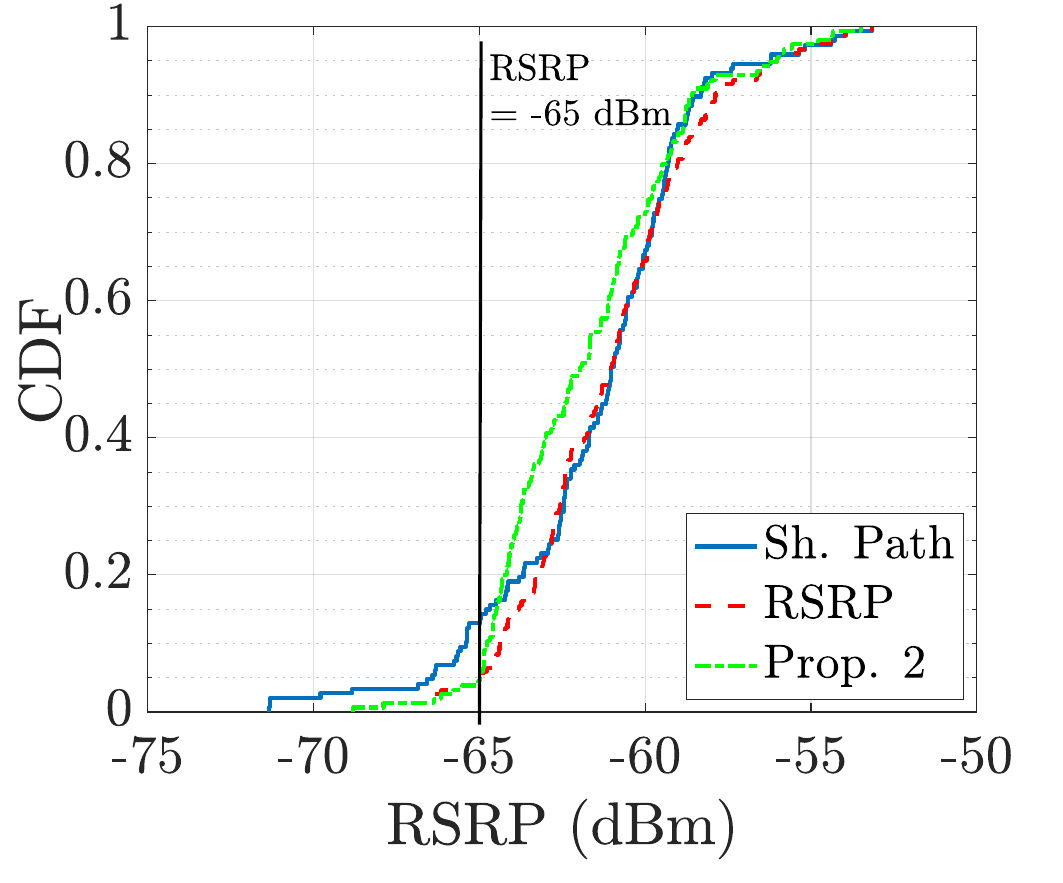}
   \caption{Comparison of RSRP's CDFs.}
    \label{fig:cdf} 
\end{figure}


\textcolor{black}{We illustrate in Fig. \ref{fig:cdf} the cumulative distribution function (CDF) of RSRP for the trajectory planning algorithms, under the same aforementioned assumptions. At the  RSRP threshold $T_{\rm th} = -65$ dBm, shown with a vertical line, the proposed method achieves the same low outage probability as the RSRP-based algorithm. Indeed, the proposed scheme's flexibility allows for reducing the ping-pong handoffs and associated signaling overhead, at the expense of a low RSRP. At RSRP values exceeding the threshold $T_{\rm th}$, the proposed method's reliability deteriorates (i.e., higher CDF) compared to the benchmarks. This is expected since our algorithm does not guarantee its optimality beyond the parameters' values it was trained with.}

\vspace{-10pt}

\section{Conclusion}
In this work, we proposed a novel energy-efficient disconnectivity and handoff aware RL-based trajectory planning and cell association algorithm for cargo-UAVs. 
Through simulations, we prove the superiority of our approach in terms of number of handoffs and disconnectivity, at the expense of a slight increase in battery consumption, when compared to baseline algorithms.


\vspace{-15pt}
\bibliographystyle{IEEEtran}  
\bibliography{references,Bibliography}

\end{document}